\documentclass[preprint, times, number, sort&compres]{elsarticle}
\usepackage{graphicx,epsfig}
\usepackage{amsmath}
\usepackage{hyperref}

\begin{document}
\title{Force correlations in molecular and stochastic dynamics}
\author[rvt]{Su Do Yi}
\author[rvt]{Beom Jun Kim\corref{cor1}}
\ead{beomjun@skku.edu}
\cortext[cor1]{Corresponding author}
\address[rvt]{Department of Physics and BK21 Physics Research Division,
Sungkyunkwan University, Suwon 440-746,
Korea}
\begin{abstract}

A molecular gas system in three dimensions is numerically studied by the energy 
conserving molecular dynamics (MD). The autocorrelation functions for 
the velocity and the force are computed and the friction coefficient is 
estimated. From the comparison with the stochastic dynamics (SD) of a 
Brownian particle, it is shown that the force correlation function 
in MD is different from the delta-function force correlation in SD 
in short time scale. However, as the measurement time scale is increased 
further, the ensemble equivalence between the microcanonical MD 
and the canonical SD is restored. We also discuss the practical implication
of the result.

\end{abstract}


\maketitle

\section{Introduction}
Since Einstein published a seminal paper on the Brownian motion in 1905, it
became one of the most well-established subjects in statistical physics. The
motion of a Brownian particle has been studied in many works theoretically and
numerically, and was extended later to L\'evy
noise, L\'evy flights, L\'evy walks, continuous time random walks, fraction
diffusion, etc. These extensions are being used to describe complex phenomena,
e.g., anomalous diffusive behaviors~\cite{Sand} or the diffusion limited growth
and aggregation mechanisms~\cite{Soko}. For physicists, the study of
Brownian motion led to a broad class of equations of motion containing
various stochastic effects. Especially, the Langevin equation~\cite{Reif} is
the most representative differential equation with the stochastic random
forces of the white noises.
Mori~\cite{Mori} derived the generalized Langevin equation
with non-Markovian noises, in which the memory kernel plays an important role.
In contrast, the original Langevin equation does not have a finite memory.

Molecular dynamics (MD)~\cite{MD} is a computer simulation method to describe
atoms, molecules, and even stars, interacting with each others in a closed
system. Since the interaction in molecular dynamics obeys classical mechanics,
the positions and momenta of particles are calculated by numerical integration
of Newtonian equations of motion. It is important to note that the total energy
of the system in MD is conserved as the system moves along the trajectory in
phase space. In this regard, the fundamental hypothesis of equal {\it a priori}
probability in statistical mechanics ensures that the MD simulation basically
generates the microcanonical ensemble.  In the computational point of view, the
number of particles in MD is often limited and far less than any real system
because of the limitation in computer capacity.  Even with this drawback,
MD simulation has been proven to be an excellent approximation for
investigation of a variety of classical quantities of real materials.

The computational limitation by the all-particle approach in the microcanonical
MD can be overcome if one adopts a different approach. Imagine that we can
conceptually divide the whole system into two subsystems: One is composed of the
degrees of freedom
we like to trace, and the other is
the environment system that plays the role of heat reservoir.
In the MD approach, we need to
integrate all particles' equations of motion. However, if it is possible
to describe interaction by environmental degrees of freedom as
stochastic random forces to the system variables, numerical integrations
become much lighter simply due to the reduction of the number of degrees
of freedom we need to trace. In this stochastic dynamics (SD) approach,
we only need to integrate equations of motion for system particles, and
effects from other environmental particles are handled as stochastic
random forces that satisfy some given statistical properties.

In the present work, we use the MD approach with total number
$N$ of particles, the volume $V$ of the system, and the total
energy $E$ fixed ($NVE$ ensemble in MD)~\cite{MD}. In this MD method,
it is to be noted that the time evolution of the system is deterministic
and only depends on initial conditions within numerical accuracy. During 
the MD simulation, one is
allowed to look at a single particle (call it a Brownian particle) and
consider every other particles as composing an environment system, applying
forces to the Brownian particle from time to time. This simple change
of view allows us to make connection between the MD and the SD approaches,
which composes the main theme of the present paper.
In equilibrium, the energy conserving
microcanonical ensemble is equivalent to the energy fluctuating canonical
ensemble in thermodynamic limit~\cite{Sali} in most situations,
with some interesting exceptions~\cite{mychoi}.
In the present context, the key question to pursue in our work is in what
condition the ensemble equivalence between the MD and the SD approaches becomes
valid, in parallel to the ensemble equivalence between the microcanonical and
the canonical ensemble in equilibrium statistical mechanics.
In more detail, we simulate the MD with the $NVE$ ensemble and
observe the motion of a Brownian particle in viewpoint of the SD.
All the combined applied forces by other particles on the Brownian particle
are interpreted as the effective stochastic forces, and the motion
of the Brownian particle is compared with that from the simple Langevin
equation with stochastic random force.

The present paper is organized as follows: In Section~\ref{sec:methods}, we describe the details of our MD simulation method and quantities to be measured.
The obtained results are shown in Section~\ref{sec:results} in comparison with the
SD approach of a Brownian particle, which is followed by the summary
in Section~\ref{sec:summary}

\section{Simulation Methods}
\label{sec:methods}

In our simulations, we use $N$ molecules of the identical mass $m$
in the presence of the truncated Lennard-Jones interaction
called the WCA (Weeks, Chandler, and Andersen) potential~\cite{WCA},
which contains only the repulsive part of the Lennard-Jones interaction:
\begin{equation}
V_{\rm WCA}(r)=
\begin{cases}
4\epsilon\left[\left( \frac{\sigma}{r}\right)^{12}-\left(\frac{\sigma}{r}\right)^{6}\right]+\epsilon,
& \mbox{for $r<r_c$},\\
0, & \mbox{otherwise},
\end{cases}
\end{equation}
where $r_c \equiv 2^{1/6} \sigma$ with the interaction length scale
$\sigma$ and the interaction strength $\epsilon$.
This representation of potential guarantees continuity of the potential
and the force at $r = r_c$, i.e.,
$V_{\rm WCA}|_{r = r_c -} = V_{\rm WCA}|_{r = r_c +} = 0$ and
$(dV_{\rm WCA}/dr)|_{r = r_c -} = (dV_{\rm WCA}/dr)|_{r = r_c +} = 0$.
We make equations of motion dimensionless by choosing
$\sigma, m$, $\epsilon$, and $\sqrt{m\sigma^2/\epsilon}$
as units of the length, the mass, the energy, and the time, respectively,
and obtain
\begin{equation} \label{eq:md}
\ddot{\mathbf{r}}_i =
48\sum_{j \neq i}^{'} \left(r_{ij}^{-14}-\frac{1}{2}r_{ij}^{-8}\right)\mathbf{r}_{ij} ,
\end{equation}
where $\sum'$ denotes that the sum is over only molecules ($j$'s) satisfying
$ r_{ij} \equiv | \mathbf{r}_{ij} (\equiv \mathbf{r}_j - \mathbf{r}_i) | < r_c$.
Because the force term on the right-hand-side of Eq.~(\ref{eq:md}) does
not depend on the velocity, we can use the following numerical integration
scheme: $v(t+\Delta t/2)=v(t-\Delta t/2)+a(t)\Delta t$ and
$x(t+\Delta t)=x(t)+v(t+\Delta t/2)\Delta t$ with the position $x$, the
velocity $v$, the acceleration $a$, and the discretized time step size
$\Delta t$. This method is called the leap-frog algorithm and
it is easy to see that this is the second-order algorithm although
it runs at the same speed as in the simple first-order Euler method~\cite{MD}.

In MD simulations, a three-dimensional cubic box of the linear size $L$ (the
volume $V=L^3$) with $L=24$ (in unit of $\sigma$) under periodic boundary
condition is used. The total number of particles is set to $N=864$
(the number density is thus fixed to $\rho = N/V = 1/16$) to make
initial positions of particles fit to face-centered cubic (fcc) structure,
in order to avoid the huge amplitude of the force which might happen
if the particles are scattered initially at random positions.
The discretized time step in numerical integration is $\Delta
t=2\times 10^{-4}$, which is small enough so that further decrease
of $\Delta t$ does not change results reported in the present paper.
The total simulation time is $2\times10^6\Delta t = 400$.
We also verify that the total energy $E$ of the system is conserved
within numerical accuracy. The total momentum should also be conserved
and can be set to zero for convenience.

\begin{figure}
\includegraphics[width=0.95\textwidth]{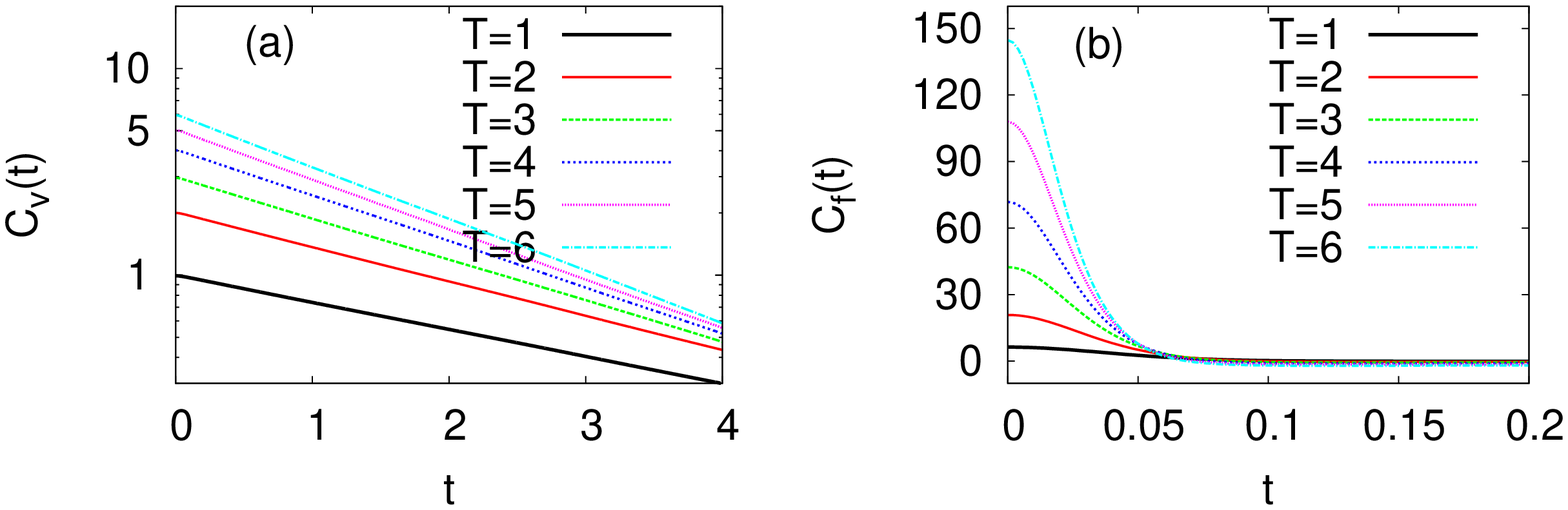}
\caption{(a) The velocity $C_v(t)$
and (b) the force $C_f(t)$ autocorrelation functions versus time $t$.
The velocity autocorrelation $C_v(t)$ decays exponentially in time
in accord with the result from the Langevin equation for a Brownian
particle. [Note that $C_v(t)$ in (a) is in the log scale].
In (b), $C_f(t)$ crosses the horizontal axis at the cutoff time $t_c$.
The friction coefficient $\gamma_v$ is computed from $C_v(t)$
via the Einstein approach, which is then compared with
the friction coefficient $\gamma_f$ from $C_f(t)$
(see text and Table~\ref{tab:gammaG}).
}

\label{fig:vf}
\end{figure}

The equipartition theorem $(1/2)\sum_{i=1}^{N}m_i v_i^2=(3/2)N k_B
T$ with the Boltzmann constant $k_B$ is used to calculate the temperature $T$.
We first generate uniform random velocities in [-1,1] and shift them to make
the total momentum (or the velocity of the center of mass) zero.
The initial temperature $T'$ is then computed from the equipartition
theorem, and  we scale the velocities according to
$\mathbf{v}_i \rightarrow \mathbf{v}_i\sqrt{T/T'}$ to tune the system
at the given temperature $T$. As time proceeds, the system approaches
equilibrium and the temperature is computed (call it $T''$)
again from the equipartition
theorem. The deviation from the input temperature $T$ is removed by
the second velocity scaling $\mathbf{v}_i \rightarrow \mathbf{v}_i\sqrt{T/T''}$.
After this procedure, we confirm that the temperature does not deviate
much from the input temperature $T$.
In our $NVE$ ensemble MD simulations,
the temperature is the control parameter and we use
$T=1,2,3,4,5$ and $6$ in units of $\epsilon/k_B$.
It is then straightforward to numerically integrate the equations of
motion~(\ref{eq:md}) and all the presented results are obtained
from the average over 100 independent runs. The key quantities we measure
during simulation is the force $\mathbf{f}_i$ [the right-hand-side
of Eq.~(\ref{eq:md})] and the velocity $\mathbf{v}_i$, which are
then used to compute the velocity and the force autocorrelations defined by
$C_v(t) \equiv \langle v_i^{(\alpha)}(t') v_i^{(\alpha)}(t'+t) \rangle$
and $C_f(t) \equiv \langle f_i^{(\alpha)}(t') f_i^{(\alpha)}(t'+t) \rangle$,
respectively, with $i = 1, 2, \cdots N$, $\alpha = x,y,z$, and $\langle \cdots \rangle$ being the average over particles ($i$), directions ($\alpha$), and time ($t'$),
after a sufficiently long equilibration time (see Fig.~\ref{fig:vf}).

Before we delve into the interpretation of our MD results, we briefly
review the Einstein approach for the stochastic Langevin equation of a Brownian
particle~\cite{Reif,Fokk}:
$m \ddot{\mathbf{r}}=-\gamma \dot{\mathbf{r}} + \mbox{\boldmath$\eta$}$,
where $m$ is the mass of the particle, $\gamma$ is the coefficient of the
viscous friction, and {\boldmath$\eta$}
is the stochastic random force assumed to be Gaussian white noise.
In the same dimensionless units as in our MD, the Langevin equation is written as
\begin{equation}
\label{eq:sd}
\ddot{\mathbf{r}}= -\gamma\dot{\mathbf{r}}
+ \mbox{\boldmath$\eta$},
\end{equation}
with $\gamma$ and $\eta$ are in units of $\sqrt{m\epsilon/\sigma^2}$
and $\epsilon/\sigma$, respectively,
and thus the noise correlation takes the form of
\begin{equation}
\label{eq:noise}
\langle \eta^{(\alpha)}(t')\eta^{(\alpha)}(t'+t) \rangle = 2 \gamma T \delta(t)
\end{equation}
for each component $\alpha = x,y,z$,
with $T$ and $t$ in units of $\epsilon/k_B$ and $\sqrt{m\sigma^2/\epsilon}$.
Following the Einstein
approach~\cite{Reif,Fokk}, it is straightforward to calculate
the velocity autocorrelation function
$C_v(t) = (k_B T/m) e^{- \gamma t/ m}$,
which is written as
\begin{equation}
\label{eq:cv}
C_v(t) = T e^{- \gamma t},
\end{equation}
in our dimensionless units.

\section{Results}
\label{sec:results}

Fig.~\ref{fig:vf}(a) obtained from our MD simulation of Eq.~(\ref{eq:md})
clearly shows that $C_v(t=0) = T$ and that $C_v(t)$ decays exponentially
in time, which are in perfect agreement with the result from
the Einstein approach in Eq.~(\ref{eq:cv}).
A simple curve fitting of $C_v(t)$
to the exponential function gives us the friction coefficient
$\gamma_v$ (we use the subscript $v$ to indicate that it is
computed from the velocity correlation),
which are tabulated in Table~\ref{tab:gammaG}.

Alternatively, the friction coefficient can also be found from 
the Green-Kubo formula for the force autocorrelation
function~\cite{Espanol,Kirk}:
\begin{equation}
\label{eq:greenkubo}
\gamma_f=\frac{1}{3 T}\int_{0}^{t_c}\langle
\mathbf{f}(0)\cdot\mathbf{f}(t)\rangle dt
\end{equation}
in dimensionless form, where 3 in the denominator
comes from the dimensionality. The cutoff time $t_c$ can
be taken large enough so that the force autocorrelation function has
arrived at plateau region~\cite{Espanol}. Lagar'kov and
Sergeev~\cite{Larg} have proposed another practical solution in which
$t_c$ is taken as the first zero of the force autocorrelation function.
We use the latter approach and compute the friction coefficient $\gamma_f$
based on Eq.~(\ref{eq:greenkubo}) and present results in Table~\ref{tab:gammaG}.
It is to be noted that the two different ways of computing the friction
coefficient, one from velocity autocorrelation and the other from force
autocorrelation, give us the identical results within numerical accuracy.
The agreement between $\gamma_v$ and $\gamma_f$ can also be interpreted
as implying the ensemble equivalence between the MD and the SD, since
$\gamma_v$ is based on the expression from the Einstein approach for the SD,
while the expression for $\gamma_f$ is based on the MD.

\begin{table}
\caption{Friction coefficients computed from the velocity autocorrelation and
the force autocorrelation functions [$\gamma_v$ and $\gamma_f$
in Eqs.~(\ref{eq:cv}) and (\ref{eq:greenkubo}), respectively]
at various temperatures $T$.
The time integration of the correlation function
$G_{MD}(t)$ for the MD in Eqs.~(\ref{eq:G}) and (\ref{eq:intG})
is also shown together with the cutoff time $t_c$ defined from
zero crossing of $C_f(t)$ in Fig.~\ref{fig:vf}(b).
The relaxation time scale $\tau$ of $G_{MD}(t)$ is also listed (see text).
}
\begin{tabular}
{c|c c c c c c c c}
\hline\hline
$T$ & $\gamma_v$ & $\gamma_f$ & $t_c$ &  $\int_{0}^{t_c}G_{MD}(t) dt$ &
$\tau$ \\
\hline
1.0 & 0.30(1) & 0.30(1)   &  0.118 & 0.97(1) &  0.048 \\
2.0 & 0.38(1) & 0.39(1)   &  0.096 & 1.00(1) &  0.038 \\
3.0 & 0.46(1) & 0.46(1)   &  0.086 & 0.97(1) &  0.033 \\
4.0 & 0.51(1) & 0.52(1)   &  0.078 & 0.98(1) &  0.029 \\
5.0 & 0.55(1) & 0.57(1)   &  0.072 & 1.00(1) &  0.026 \\
6.0 & 0.58(1) & 0.60(1)   &  0.068 & 1.00(1) &  0.025 \\

\hline\hline
\end{tabular}
\label{tab:gammaG}
\end{table}

In order to check the equivalence between the MD and the SD in more detail,
we next study the force autocorrelation function.
For this, we first start from Eq.~(\ref{eq:sd}) and write
\begin{eqnarray}
\label{eq:sdff}
\langle f(0)f(t) \rangle & = & \langle [-\gamma v(0) +\eta(0)][-\gamma
v(t)+\eta(t)] \rangle \nonumber \\
& = & \gamma^2 \langle v(0) v(t) \rangle + \langle \eta(0)\eta(t) \rangle
\end{eqnarray}
for one direction (we have skipped the index $\alpha$ for direction for
brevity), and the time $t$ is measured after equilibration so that the
correlation is invariant under time translation.
We define another correlation function $G(t)$ as
\begin{equation}
\label{eq:G}
G(t) \equiv \frac{\langle \eta(0)\eta(t) \rangle}{\gamma T}=
\frac{\langle f(0)f(t) \rangle-\gamma^2\langle v(0) v(t) \rangle}{\gamma T} .
\end{equation}
In MD, we know all velocities and forces at each time, which are used
to compute $G(t)$ in Eq.~(\ref{eq:G})
[we call it $G_{MD}(t)$],
combined with the friction coefficient computed above (Table~\ref{tab:gammaG}).
On the other hand, the corresponding correlation function $G_{SD}(t)$
for SD is written as $G_{SD}(t) = 2 \delta(t)$ from
Eqs.~(\ref{eq:noise}) and (\ref{eq:G}).

\begin{figure}
\includegraphics[width=0.95\textwidth]{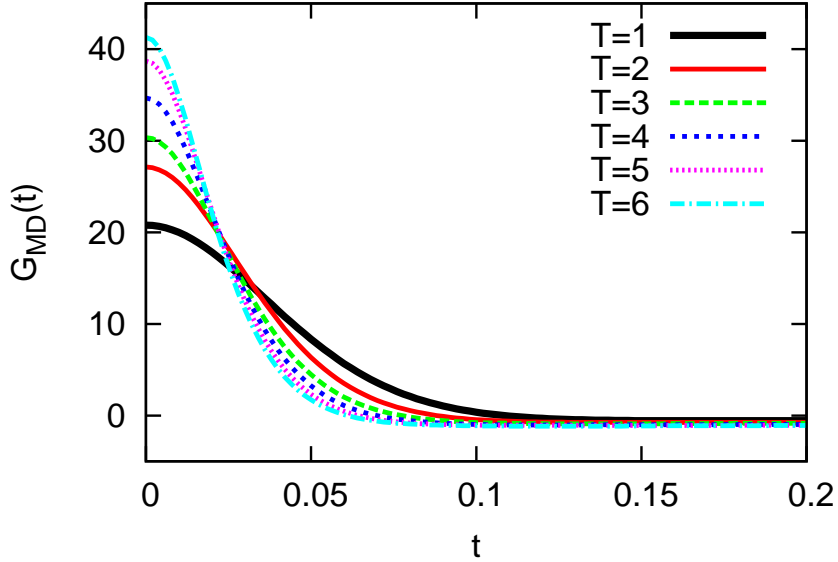}
\caption{The correlation function $G_{MD}(t)$ versus time $t$ in
Eq.~(\ref{eq:G}) calculated for the MD simulation. $G_{MD}(t)$ decays in
time with finite relaxation time (see text and Table~\ref{tab:gammaG}), which differs from $G_{SD}(t) = 2\delta(t)$.
}
\label{fig:G}
\end{figure}

Fig.~\ref{fig:G} shows $G_{MD}(t)$ at different temperatures.
The relaxation time scale $\tau$ for $G_{MD}(t)$ can be
estimated by using the method in Ref.~\cite{tau} with the time
integration up to $t_c$ (see Eqs.~(8)-(10) in Ref.~\cite{tau}).
The resulting values of $\tau$ at various temperatures are tabulated
in Table~\ref{tab:gammaG}.
If the observation time scale is much larger
than the relaxation time scale $\tau$ of MD, we expect one can approximate
$G_{MD}(t) \approx G_{SD}(t) = 2 \delta(t)$. To confirm the equivalence
between MD and SD in such a long-time scale, $G_{MD}(t)$ needs to satisfy
\begin{equation}
\label{eq:intG}
\int_{0}^{t_c} G_{MD}(t) dt = \int_{0}^{t_c} G_{SD}(t) dt =
2 \int_{0}^{t_c} \delta(t)dt = 1,
\end{equation}
where the last equality comes from the evenness of the delta function in
time, and in the same spirit as in Eq.~(\ref{eq:greenkubo}) we have used
$t_c$ (see Table~\ref{tab:gammaG}) as the cutoff of
the integration in Eq.~(\ref{eq:intG}).
We find that the equivalence between the MD and the SD
is convincingly borne out as listed in Table~\ref{tab:gammaG}, where
$\int_0^{t_c} G_{MD}(t) dt \approx 1$ at all temperatures, as expected.

\begin{figure}
\includegraphics[width=0.95\textwidth]{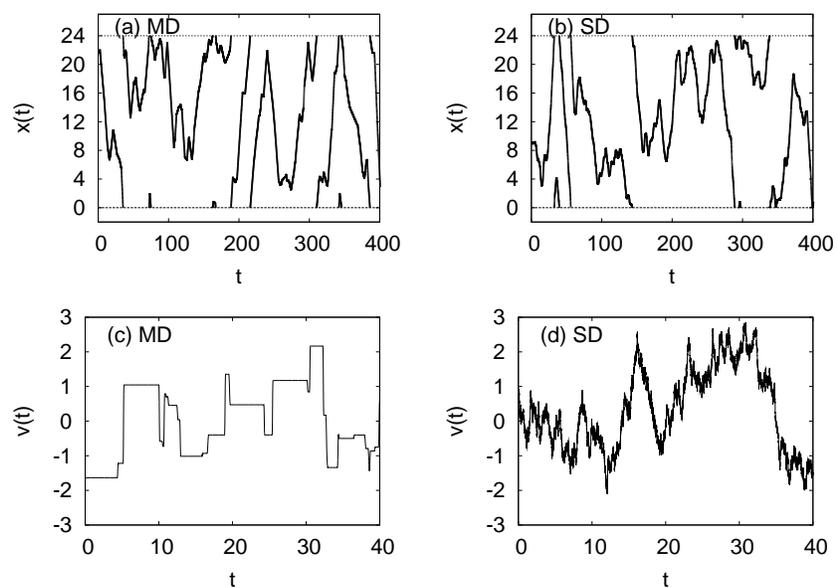}
\caption{The position of a particle in $x$ direction versus time $t$
is shown for (a) the MD and (b) the SD in a long-time scale of $O(10^2)$.
(c) and (d) display the
velocity $v(t)$ for the MD and the SD, respectively, in
a shorter-time scale of $O(10)$. When the observation time scale is
long enough, trajectories from the MD and the SD are qualitatively
the same [see (a) and (b)]. On the other hand, in a shorter-time scale,
they behave very differently [see (c) and (d)].}
\label{fig:xvt}
\end{figure}

In comparison to the MD, the use of SD has a significant benefit in practical
point of view due to the small number of degrees of freedom to integrate.
The key question to answer is then what is the condition for the equivalence
between the MD and the SD. We find above that the observation time scale
must be sufficiently large so that the correlation function $G_{MD}(t)$ can
be approximated as a delta function as in $G_{SD}(t)$ to see the
consistency between the SD and the MD: if we are only
interested in long-time behavior, it is reasonable to use the SD instead
of the MD. In other words, although the motion of the particle in the MD
in short-time scale is very different from that of the simple Langevin
dynamics, the two cannot be distinguished in long-time scale.

\begin{figure}
\includegraphics[width=0.95\textwidth]{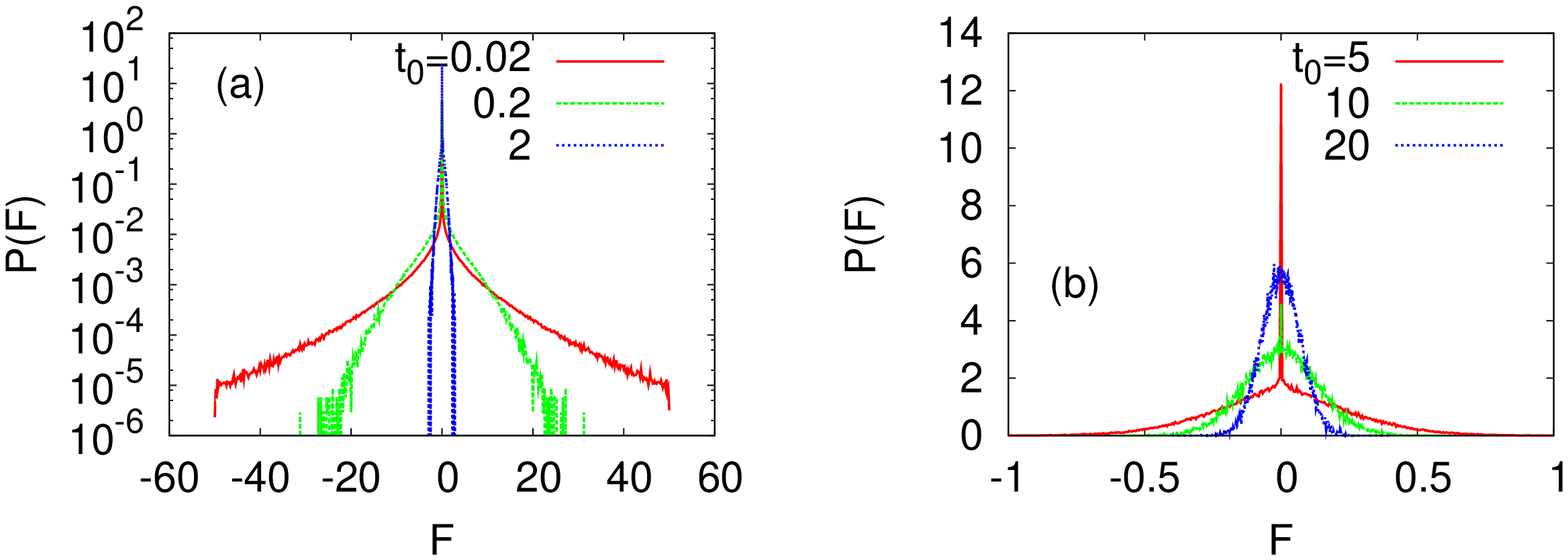}
\caption{The probability density function $P(F)$ for the time averaged force
$F$ for the time interval $t_0$. As $t_0$ is increased [from (a) to (b)],
$P(F)$ approaches the Gaussian form, again implying the ensemble equivalence
between the MD and the SD. }
\label{fig:force}
\end{figure}

Fig.~\ref{fig:xvt} displays trajectories $x(t)$ for (a) the MD and (b) the SD
at the same temperature $T=1$ in a long-time scale. In our SD simulation,
we integrate the equation of motion for a Brownian particle by using the
Runge-Kutta second-order algorithm with the same discrete time step
$\Delta = 2\times 10^{-4}$ in dimensionless time unit.
As expected, the two
trajectories look qualitatively the same. In contrast, $v(t)$ in
a shorter-time scale for (c) the MD and (d) the SD look quite different:
for MD with the WCA potential, particles interact only when the distance
between them is smaller than $r_c$, and thus $v(t)$ changes in time
in a step-like fashion. The ensemble equivalence between the MD and
the SD in a long-time scale displayed in Fig.~\ref{fig:xvt} can also
be seen in the probability density function $P(F)$ for the time-averaged
force $F$ in Fig.~\ref{fig:force}, where $F(t) \equiv (1/t_0)\int_{t}^{t+t_0}
f(t')dt'$. As the measurement time scale $t_0$ becomes larger, $P(F)$
approaches the Gaussian distribution, again revealing the ensemble equivalence
between the MD and the SD in a long-time scale. The similar ensemble equivalence
has also been discussed as the mass ratio between the Brownian particle
and the environment particle is varied~\cite{EOK}.

\section{Summary}
\label{sec:summary}

In summary, we have studied the gas system with the WCA potential
within the MD approach and compared
the results with the SD based on a simple Langevin equation.
The velocity and the force autocorrelation functions have been
computed and a good agreement has been observed in the friction coefficient
calculated independently from each correlation function.
It has been revealed that as the observation time scale becomes much
larger than the relevant relaxation time scale of correlation
function, the ensemble equivalence between the microcanonical MD
and the canonical SD approaches is established.
It is to be noted that the present study and the results are limited by the
relatively low particle density. As the particle density is increased, the
system will become liquid-like. In this liquid regime, the autocorrelations
become more extended in time, and the generalized Langevin formulation with
memory needs to be used.

\section*{Acknowledgements}
This work was supported by the National Research Foundation of Korea (NRF) grant
funded by the Korea government (MEST) (No. 2011-0015731).

\section*{References}
\bibliographystyle{elsarticle-num}

\end{document}